# A New Tool for Seismology-the Cumulative Spectral Power

**Randall D. Peters**
**Department of Physics**
**1400 Coleman Ave.**
**Mercer University**
**Macon, Georgia 31207**

**ABSTRACT**
The power spectral density (PSD) function is commonly used to specify seismometer performance. It is derived from the FFT of acceleration and correction is made for the transfer function of the instrument that generated the data. As with any such spectrum of density (`per Hz') type, the noise inherent to a PSD is large. This article illustrates the value of a function that is derived from the PSD and for which the influence of noise is significantly reduced. Called the cumulative spectral power (CSP), it is obtained from the PSD through the noise-reducing process of integration. The maximum of the CSP (corresponding to the longest graphed value of the period) provides a means for estimating the total vibrational power of the earth. The present author has significantly simplified the process of PSD generation. Thus routine graphing is straightforward-of first the FFT, followed by the generation of both a PSD and its associated CSP. The unique properties of the CSP make it valuable for the study of a variety of earth dynamics. For example, the strking simplicity of a CSP graph generated from a record containing a strong teleseismic earthquake is undoubtedly important to the development and refinement of any viable theory of earthquake dynamics.

**Background**
The PSD and the CSP are related in the same manner as a pair of well known math functions used to quantify probabilities. The probability density function (pdf) has similarities to the PSD. In the case of the normal distribution, the pdf is the `bell curve' familiar to many students; i.e., the normalized form of Gauss' function. For the area under the curve equal to unity, the function is given by

$$f(x) = \frac{1}{(2\pi)^{1/2}\sigma} \exp(-[(x-\mu)/\sigma]^2/2) \qquad (1)$$

Mathematicians also use the cumulative probabilty function (cpf), designated by $F(x)$, and for which

$$\frac{d}{dx} F(x) = f(x) \qquad (2)$$

These functions are illustrated in Fig. 1, where the curves were generated with a mean value $\mu = 1$ and a standard deviation $\sigma = 0.3$.

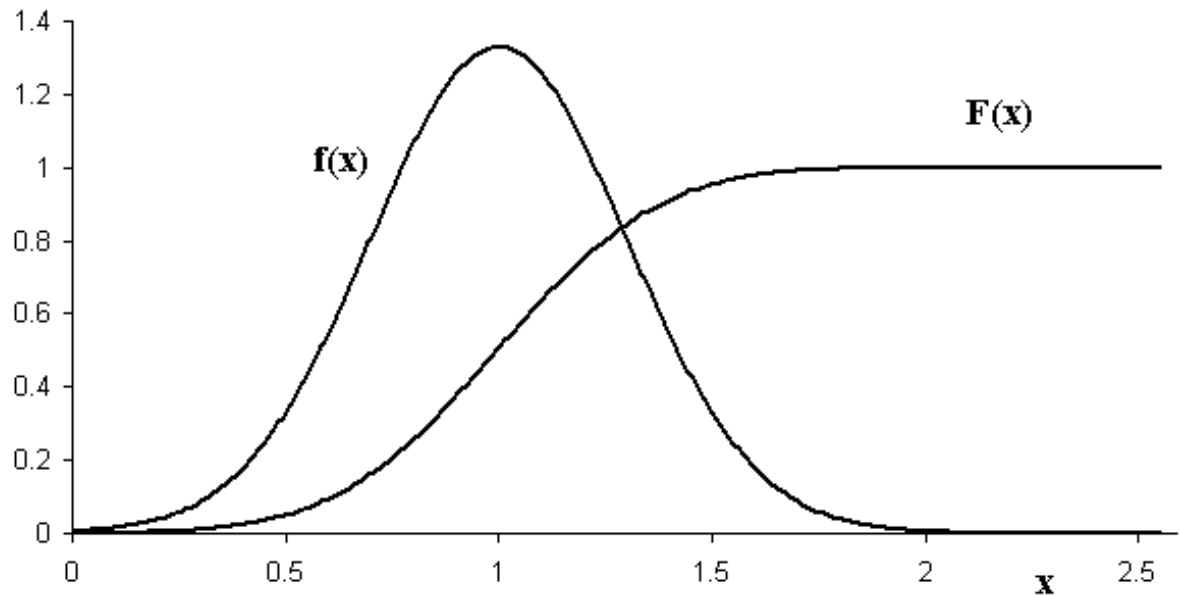

**Figure 1.** Illustration of the probability density function f(x) and the cumulative probability function F(x).

So that the PSD and CSP may be better understood (especially in terms of the density feature of the PSD, for which there is much confusion), the functions graphed in Fig. 1 are now described according to their meaning. The function f(x) specifies the probability that the magnitude of a randomly drawn variate x, governed by the normal distribution, will be located within the interval x to x + dx. Since the variate must have a value that is contained by the interval spanning plus and minus infinity, we see that the integral over f(x) must be equal to unity.

From Eq.(2) we obtain the following result for F(x):

$$F(x) = \int_{-\infty}^{x} f(x)dx \qquad (3)$$

For a given value of x, the value of F(x) is seen to be the probability that a randomly drawn variate of the distribution will be less than x.

Although probability functions are inherently smooth, noise can be nevertheless arbitrarily introduced into f(x) to show that F(x) is relatively immune to the noise, as shown in Fig. 2.

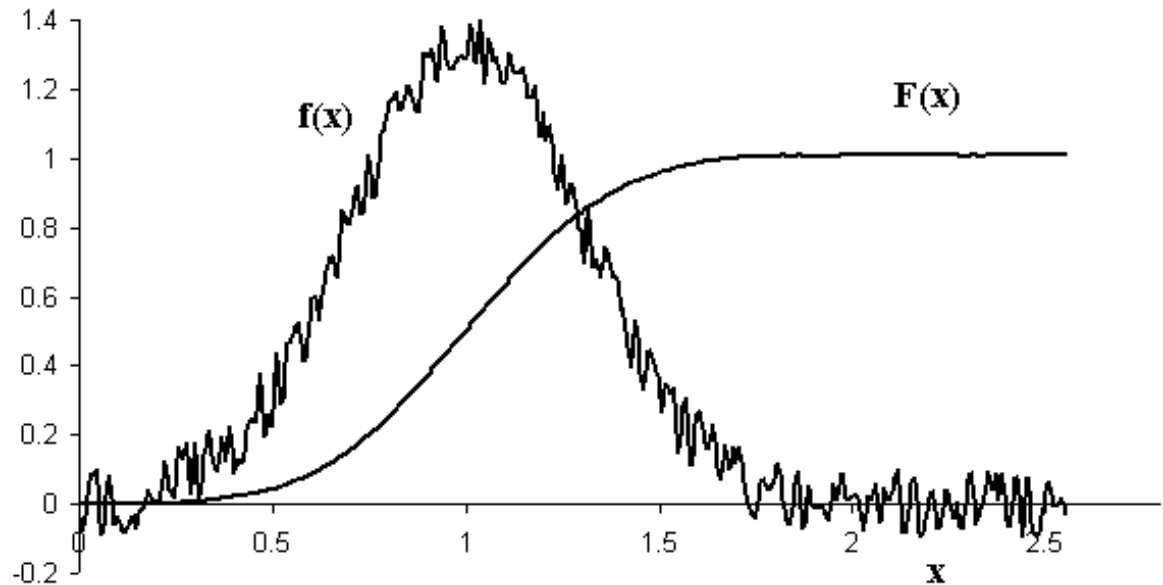

**Figure 2.** Illustration of the improvement in signal to noise ratio of a cumulative function as compared to the density function to which it is related.

We now generalize this discussion of f(x) and F(x) to the case of the PSD and the CSP, both of which can be plotted either versus frequency ν or versus period T. Because of the well-known study of earth noise in which `acceleration power' is graphed versus period [1], we choose to express both functions in terms of T. Whereas the function F(x) has the lower bound 0 and the upper bound 1, the CSP is unbounded, although it is always greater than 0. At the shortest periods of earth motion to which a seismometer responds, much if not most of the power is due to localized, man-made activities. We thus restrict our attention to periods longer than about 2 s. There is also a restriction placed on the maximum allowed value of T. The longest periods allowed in graphing the CSP are determined by (i) the duration of the data-record, and (ii) the corner frequency of a digital filter used in the data processing. For the software presently employed, there is a minimum frequency limit of 0.5 mHz (T = 2000 s) placed on the high-pass digital filter through which the VolksMeter [2] data passes before doing the FFT. This high-pass operation is used in lieu of other means for removing the adverse effects of secular trends in the data (end point differences causing problems, related to the Gibbs effect). A popular alternative to high-pass filtering is to `condition' the data with a Hanning window. Although other window types are also possible, the `cosine taper' (Hanning) is apparently a seismology standard [3]. For those familiar with optical systems, the Hanning window is to the 1-dimensional FFT-world what an `apodizing' function is to the 2-dimensional FFT-world. Apodization removes the `sharp edges' of diffraction-type from images.

By analogy with the pdf, the function PSD(T) gives the amount of earth power in watts per kg operating in the period interval between T and T+dT. In the frequency-domain of the electrical engineering world, where functions of this type orginated, the power spectral density is described by units of watts in the frequency interval from ν to ν + dν. For example, it specifies the amount of Joules per second in the frequency-bin of width dν, starting at ν, that is converted to Joule heat in a resistive load (usually stated as W/Hz). An alternative engineering form is one in which the noise voltage is specified as Volts per root Hz. For a 1-ohm resistive load, the power $V^2/R \rightarrow V^2$ and so the connection between the voltage and power specifications is for this case obvious.

**Terminology problems**

Much confusion in the seismology world has resulted from `bad science' in the form of a terminology misnomer propagated by engineers. This author views adherence to the following practice as inexcusable: The PSD ``...describes how the power of a signal or time series is distributed with frequency. Here power can be the actual physical power, or more often, for convenience with **abstract signals, can be defined as the squared value of the signal**, that is, as the actual power if the signal was a voltage applied to a 1-ohm load [4]."

There is no reason to view the acceleration response of a seismometer as an `abstract' signal. The physics of this response has been understood from the time of Isaac Newton, at least to the extent that linear (harmonic oscillator) approximations are valid. We postpone the discussion of this physics to a later section.

**Graphing subtleties**

In addition to the confusion that results from calling `power' every-thing whose FFT components can be squared and then plotted, there is the additional challenge of properly accounting for how the density specification of the PSD depends on the nature of the plotting methodology.

Unlike the probability functions, which usually are adequately specified by means of a linear-scale for the abscissa, the PSD and the CSP generally require a scale that is logarithmic. This choice is necessary because the graphs normally encompass a total range of three or more decades of period (or frequency). Without `compression' that results from the log-scale, the utility of such graphs would be severely limited. For reason of the large range of powers encompassed as a function of period, the ordinate values are also plotted with a log-scale in the form of the decibel.

To understand the influence of log-scale compression on a density function such as the PSD, it is necessary to understand what is being plotted. The PSD is derived from FFT output involving a sequence of N components, where $N = 2^n$, with n being an integer [5]. The N components are distributed as a `comb' of frequencies, with equi-spaced `teeth' separated from each other by $\Delta f = 2 f_N / N$, which involves the Nyquist (maximum) frequency $f_N = f_s/2 - f_s/N$, where $f_s$ is the rate at which data is collected; i.e. the number of samples per s. Each of the FFT components has the same weighting factor in its contribution to the power, and the integral over frequency of the PSD (discrete approximation given by the sum of the components times $\Delta f$) yields the total (average) power. Parseval's theorem states that this must be the same power as obtained from the integral of the inverse transform in the time domain.

When the PSD is represented using a log-scale for the frequency, compression causes the spacing between points to no longer be constant as is true for the linear scale. When the discrete data points of the set are pictured as in the Fig. 3 illustration; it is immediately clear that the density specification (weighting factor) must be altered.

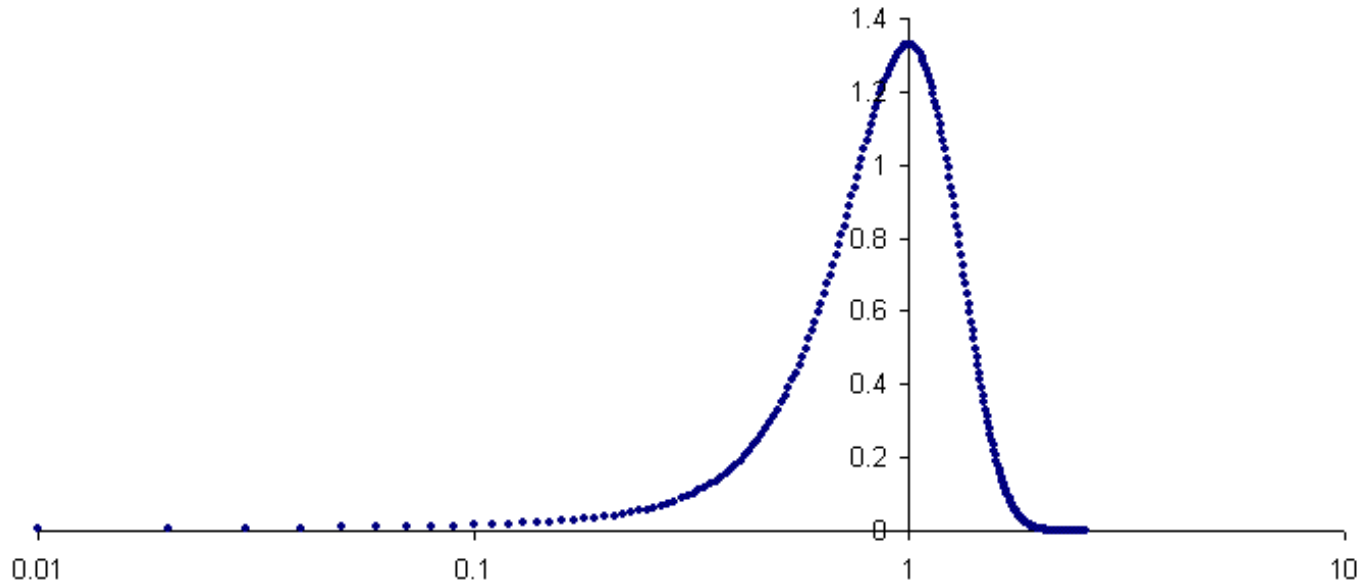

**Figure 3.** Illustration of how the probability density function of normal distribution type is altered when a log-scale is used for the abscissa.

The shape distortion of Fig. 3 is dramatic, and the number of points per decade increases as x increases. For example, the number of points in the interval from 0.01 to 0.1 is an order of magnitude smaller than the number of points in the interval from 0.1 to 1.

The function graphed in Fig. 3 is identical to the f(x) of Fig. 1, except that it has been plotted on a log-scale rather than a linear scale. If this were a PSD function, as opposed to the probability function shown, then it is clear that the density specification (the weighting factor per Hz, or per decade, or per octave, or per fraction of decade or octave, or per `whatever') must be frequency dependent for a PSD function graphed with a log-scale. For the total power to come out right, the `per whatever' must be increased as the frequency increases (proportional to f).

Another way to understand this frequency dependence of the density specification is to look formally at the mathematics. When using the log-scale, the following equality must be satisfied as a condition required for energy conservation:

$$\int f(\log x)\, d(\log x) = \int f(x)\, dx \qquad (4)$$

Since $d(\log x) = dx/x$, we see that as x increases, dx must be made smaller if the density specification is to remain constant. Since dx cannot be altered there must be a change to the density specification when using the log-scale. (Note: graphs we will discuss are generated with base 10 logarithms rather than the natural log as used in stating $d(\log x) = dx/x$; the two differ only by the constant 2.30258 or its reciprocal.)

To plot versus linear T also gives a different density specification than linear $\nu$. This is easily seen from the mathematics, since $T = 1/\nu$ and therefore $dT = -d\nu/\nu^2$. Figure 4 is a plot of the probability density function versus 1/x; it illustrates `compression' even though the scale of 1/x is itself linear.

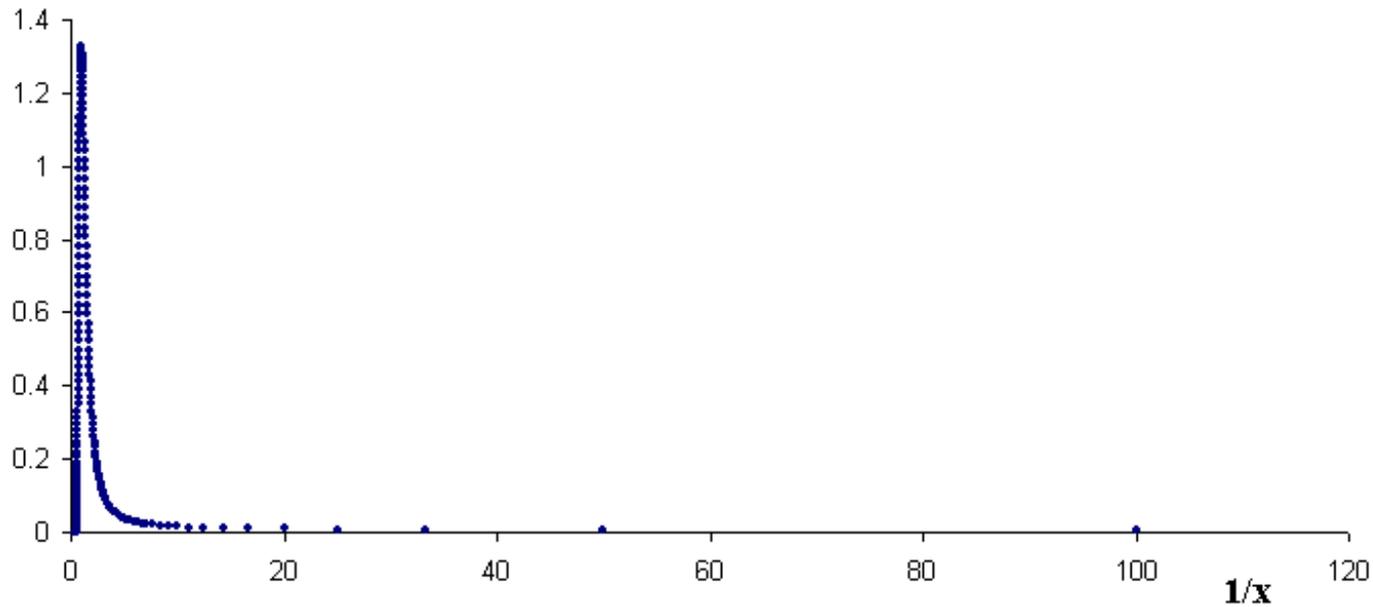

**Figure 4** Plot of the probability density function versus 1/x using a linear scale. Compression at small values of 1/x makes this graph virtually worthless.

If we re-plot Fig. 4 using log(1/x), the result is seen from Fig. 5 to be the same as Fig. 3 except for the `reversal' of direction. In other words, the density specification is the same for T and ν when using the log-scale. All one need do to convert from T to ν or vice-versa is to simply compute the reciprocal of the abscissa value.

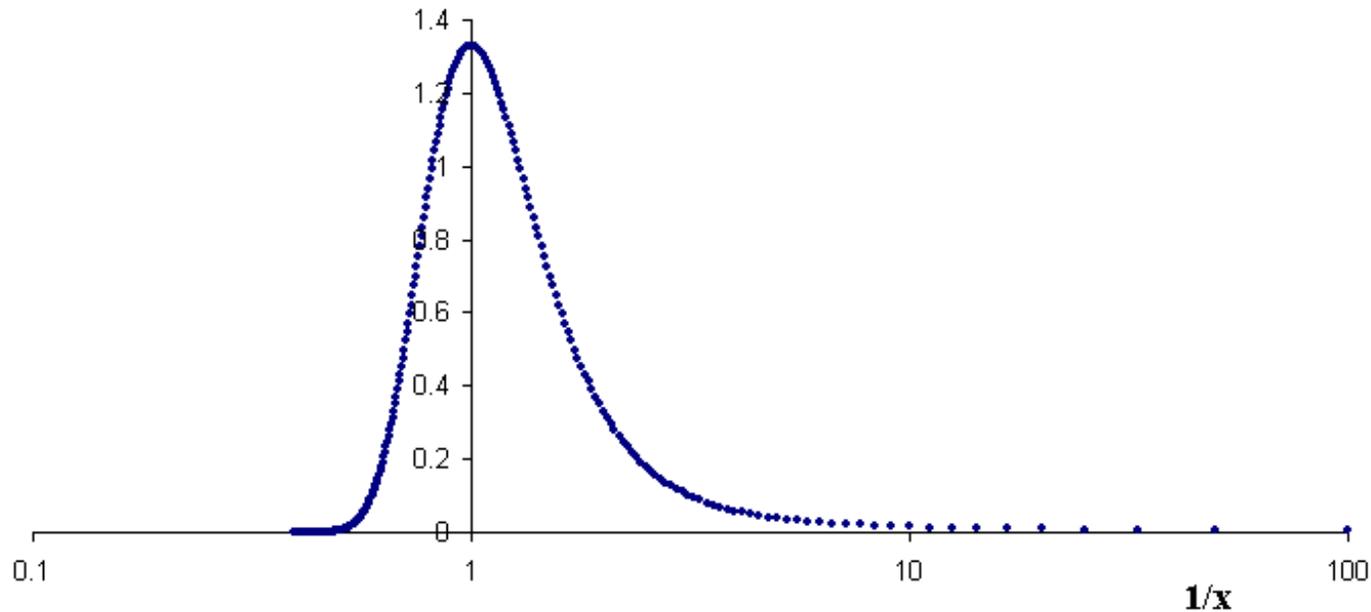

**Figure 5.** Same as Fig. 4 but using log-scale for 1/x.

## 1  Mechanical (harmonic) Oscillator Theory

The response of a pendulum is governed (linear approximation) by the following equation of motion:

$$\ddot{\theta} + \frac{\omega_0}{Q}\dot{\theta} + \omega_0^2 \theta = -\frac{\omega_0^2}{g} a(t) \qquad (5)$$

where θ is the angular displacement and a(t) is ground acceleration, related simply to ground displacement $A_G$ for steady state motion at angular frequency ω = 2π ν by a(t) = $\omega^2 A_G$. This equation approximates actual damping (largely internal [6]) through the linear viscous form (term involving the quality factor Q). For seismic purposes, Q is usually not much different from the value of 1/2 corresponding to critical damping. The natural free-decay (without imposed damping) period of the pendulum is given by T = $2\pi/\omega_0$.

To specify displacement x in nanometers of a point on the physical pendulum, one multiplies the sine of the radian angular displacement θ by the distance from the axis to the center of percussion $L_P$. Then for small θ and acceleration at the angular frequency ω, the equation of motion becomes

$$\ddot{x} + \frac{\omega_0}{Q}\dot{x} + \omega_0^2 x = -\frac{L_P \omega_0^2}{g} a(t) = \frac{L_P \omega_0^2 \omega^2}{g} A_G \qquad (6)$$

Not just the pendulum, but virtually all conventional seismometers are governed by an equation of motion that is very similar to Eq.(6). By means of force feedback (including force `balance') many conventional instruments tailor the response as a means for setting both the values of Q and $\omega_0$. The actual frequency dependence of the instrument response is usually specified in terms of pole/zero values, for which Eq.(6) is overly-simplified. The difference between Eq.(6) and those using pole/zero specifications is thought by this author to be less important than the difference introduced by nonlinear (complex) properties of real world instruments that cannot be properly modelled by any of the linear equations employed [7].

The steady state solution to Eq.(6) is

$$\frac{x_0}{\omega^2 A_G} = \frac{L_P}{g} \frac{\omega_0^2}{[(\omega_0^2 - \omega^2)^2 + \omega_0^2 \omega^2 / Q^2]^{1/2}} = \frac{L_P}{g} T_F \qquad (7)$$

where $x_0$ is the amplitude of the pendulum displacement at angular frequency $\omega = 2\pi \nu$, and $T_F$ is the transfer function of the instrument, an example shown in Fig. 6.

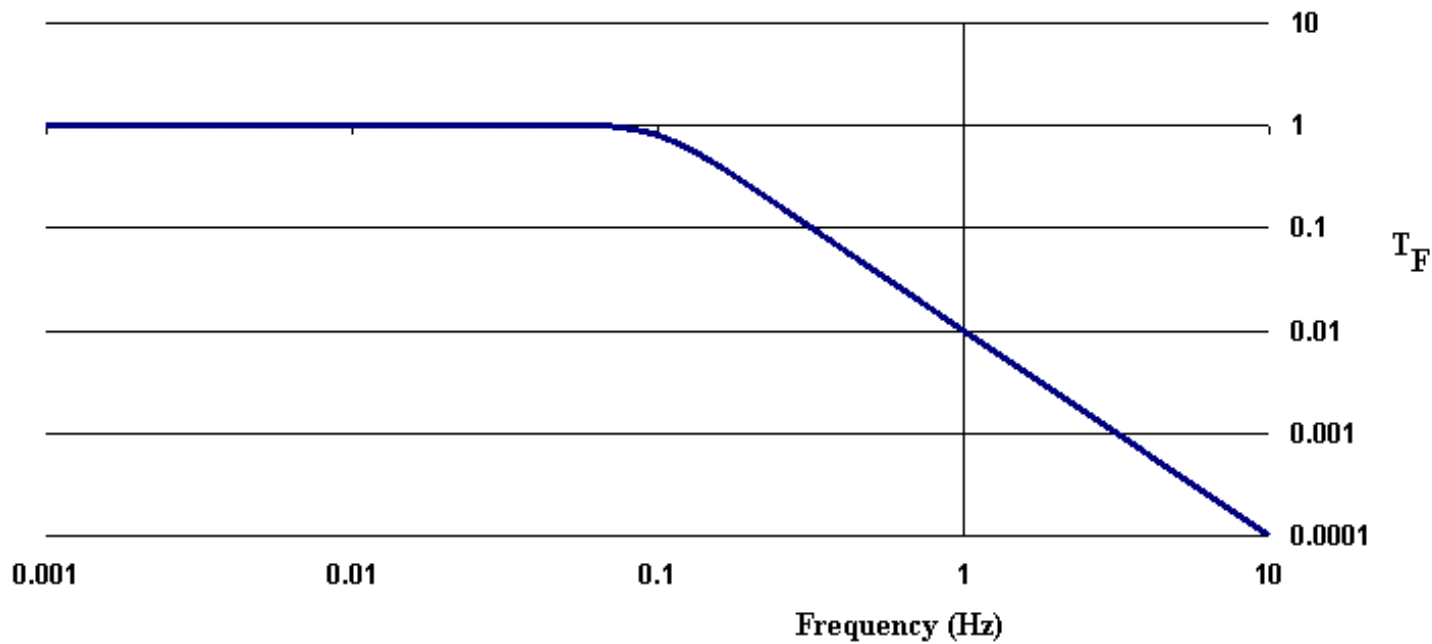

**Figure 6.** Transfer function ($T_F$, acceleration response) for an instrument with a natural period of 10 s.

It is instructive to look at Eq.(6) in the case of a simple pendulum; i.e., one in which $L_P \to l$ is simply the length of the string supporting a point mass. For this simple pendulum $\omega_0^2 = g/l$. In the limit of $\omega \ll \omega_0$ the pendulum displacement is equal to

$$x_0 = \frac{\omega^2}{\omega_0^2} A_G = \frac{l}{g} \omega^2 A_G = \frac{l}{g} \times \text{acceleration} \quad , \quad \omega \ll \omega_0 \quad (8)$$

In other words, for earth acceleration at frequency lower than the natural frequency of the instrument, the angular deflection is just equal to the acceleration divided by the earth's field; i.e., $g = 9.8$ m/s$^2$. It is important to understand that the pendulum also responds to tilt, which corresponds simply to the component of the earth's field that is perpendicular to the vertical axis of the case of the instrument. The difference between tilt response and response due to earth harmonic motion on the power spectral density is not treated here. The interested reader is referred to [8].

**Power**

Every sensor, no matter the type, responds to the stimulus exciting it according to one thing only-energy transfer (the time rate of energy being the power). Consider a mass element of the earth that oscillates in steady state. As shown above in Eq.(8), the amplitude of the acceleration measured by a seismometer observing the motion, is given by $a_0 = g\theta_0$, for frequencies of oscillation below the natural frequency of the instrument. Since the velocity of the mass is simply its acceleration divided by the angular frequency, one obtains the following expression for the total oscillatory energy of this piece of earth mass

$$E = \frac{1}{2} m v_0^2 = \frac{1}{2} m \frac{g^2 \theta_0^2}{\omega^2} \quad (9)$$

Dividing this energy by the period $T = 2\pi/\omega$ gives a value of oscillatory power. A correction to this value is provided for the following reason. Power balance involving the genertion of heat is central to our discussion. If the earth could oscillate without damping, then no external energy input would be needed to keep it oscillating. Similarly, if a seismometer were to operate without damping, no input of energy from the earth would be necessary to keep it moving. Of course damping is part of the dynamics of both systems. The damping of the instrument that monitors the earth is easy to describe in terms of the balance between power in and energy out per second in the form of heat. The quality factor is defined as $Q = 2\pi E/|\Delta E|$ where E is the total energy and $|\Delta E|$ is the loss of energy per cycle. When operating with crtical damping $Q = 1/2$, we see that the power that is converted to heat is a factor of $4\pi$ greater than the value that would be estimated on the basis of Eq.(9). Thus the following expression is used to define the specific power (power per unit mass):

$$P = \frac{g^2 \theta_0^2}{\omega} \quad , \quad \omega < \omega_0 \quad (10)$$

A completely proper accounting for the power would have to consider every term in the frequency dependence of the Q of the earth, which is viewed as a daunting task. For persistent oscillations (fairly high Q eigenmodes) Eq. (10) is insufficient, but for most of the broadband noise of the earth it is a reasonable starting point for understanding how the oscillatory power of the earth is distributed among the frequencies allowed by its density of states.

**Power Spectral Density PSD**

The expression of Eq.(10) is valid only if the indicated condition is met. The general result is obtained by correcting $\theta_0$ for the influence of instrument response; i.e., factor in the transfer function corresponding to Fig. 6. The frequency dependence of this specific power (units of watts per kg) is expressed using the decibel (involving log to the base 10). In the following equation, account has been given to both the transfer function $T_F$ and the pendulum's calibration constant $C_C$ (units of counts per radian).

$$P = 10 \log\left(\frac{g^2 \text{FFT}_{mag}^2}{\omega C_c^2 T_F^2}\right) + 10 \log\left(\frac{f_T}{2 F_N/N}\right) \quad (11)$$

where the units associated with the argument of the first log term are inconsequential because the reference for the dB specification is 1 $m^2/s^3$. All the parameters are dimensionless with the exception of g with units $m/s^2$ and $\omega$ with 1/s.

Here $FFT_{mag}^2 = 2(FFT_{real}^2 + FFT_{imaginary}^2)$ where real and imaginary refer to the complex components of the FFT of the measured values of pendulum angular deflection. The factor of two is necessary to accommodate negative frequencies which are normally not explicitly considered. Each FFT component corresponds to a specific frequency $f_T$ and the term $2 F_N/N$ is the minimum (lowest) frequency of the FFT set, seen to be equal to the Nyquist frequency divided by one half the total number of terms used in calculating the FFT. The second log term involving this ratio of frequencies is necessary when using the log-scale of either frequency or period, because of the compression described above in detail. In the example PSD graphs that follow, based on data from a fully-digital VolksMeter seismograph, the corner frequency of the transfer function is 0.918 Hz = $\omega_0/2\pi$, and $C_c = 2.5 \times 10^9$ counts/rad.

**Proper units of the PSD**
The only acceptable units of power spectral density involving actual power rather than the illegitimate `power' of abstract signals mentioned earlier - are $m^2/s^3$ per `density-unit', where the `density-unit' used in the material that follows is one-seventh decade. This density-unit is also the `bin-width' that was employed by Berger et al [1]. Another common density-unit is `Hz'.

**Erroneous labeling of the PSD**
It is now shown that seismologists work with bonafide (legitimate) PSD's, even though the units they employ are wrong. It appears that serendipity figured into the standard that developed years ago. As frequently stated to students of physics, if the units are not right, then something must be wrong.

The most common set of erroneous units are $m^2/s^4/Hz$; which should instead be $m^2/s^3/Hz$. Evidently, the label evolved from the illegitimate practice of simply squaring FFT components and graphing them. Bear in mind that the (specific) power is really watts per kg per FFT component, which equals $m^2/s^3$ per FFT component. A frequency bin-width of 1 Hz is not the same as $\Delta f = f_{min} = 2 F_N/N$ that separates adjacent spectral components of the FFT. When the FFT-squared values are plotted versus log-scale of either frequency or (more commonly) period; account must be taken for the compression discussed earlier. As noted, compression is included in Eq.(11). Moreover, by using properties of the logarithm, the equation can be rewritten as

$$P = 10 \log\left( \frac{g^2 \, FFT_{mag}^2 \, N}{4\pi \, C_c^2 \, T_f^2 \, F_N} \right) \quad , \quad \text{referenced to } m^2/s^3 \qquad (12)$$

In other words, following the historical practice and choosing to use a log-scale generates the true PSD by accident, but the units implied by that practice are incorrect, because compression has been ignored.

**Example PSD's**
Shown in Fig. 7 are some example PSD's. The red graph was generated from a record that included the indicated earthquake. The duration of the record responsible for the blue graph (`background') was the same and collected at nearly the same time but shifted enough for none of the teleseism waves to be obvious.

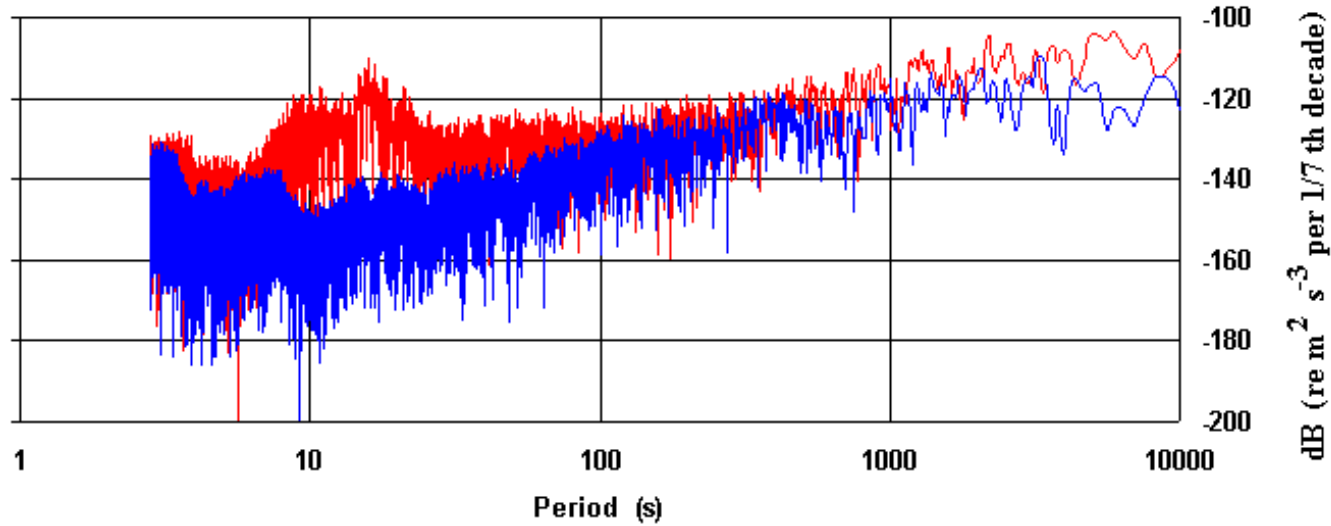

**Figure 7.** Example power spectral density graphs generated from data collected with a VolksMeter seismograph located in Macon, Georgia USA.

**Cumulative Spectral Power**
The noise that is typical of raw spectra and PSD's is quite obvious in the two cases shown in Fig. 7. As was stated earlier, such noise becomes much less significant when one works with the function that is based on the integral of the PSD; i.e., the Cumulative Spectral Power (CSP), some examples of which are shown in Fig. 8.

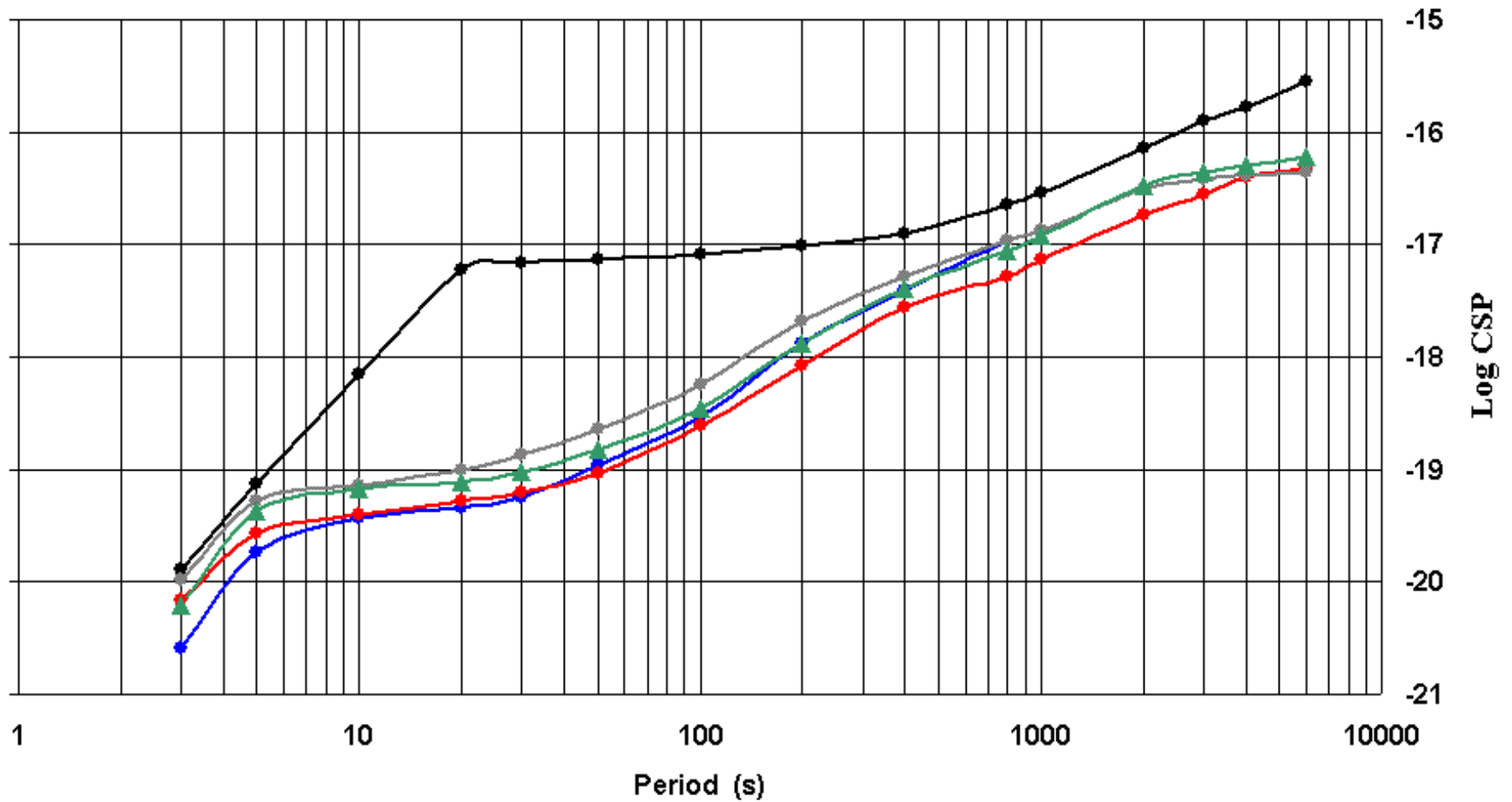

**Figure 8.** Examples of the Cumulative Spectral Power. The black curve was generated by numerical integration over the red curve of Fig. 7.

Instead of using dB's as in the case of the PSD, the CSP's of Fig. 8 have all been plotted using a simple log-scale. These curves were generated with a small number of period- values (15) as compared to the number of points of the FFT (32,768) used to generate their associated CSP. The sum was performed using Excel after exporting the txt (comma separated variable) form of the FFT.

The value of the function for a given period-value is obtained by summing all the FFT components from the starting period (2.8 s) up to and including the component that corresponds most closely to the given period-value and then multiplying that sum by $\Delta f \; = \; 2 \, F_N/N$.

The smoothing that results from the integration process that defines the CSP is dramatic, as can be seen by comparing the earthquake curve that is common to figures 8 and

7. Also stunning is the simplicity of that earthquake CSP as compared to its PSD. Data representations of physics that have become famous through the years have always been those that interject simplicity into otherwise complex problems. The simplicity of the black curve of Fig. 8 suggests that earth motions derived from large earthquakes might be modelled more simply and thus effectively than most have wanted to believe.

**Total Earth Power**
The CSP is derived from specific power (W/kg) of the PSD. If the FFT used to generate the PSD is properly normalized, then the ordinate values of the Fig. 8 graph take on absolute meaning as opposed to the more typical view of spectra in terms of arbitrary units (for relative comparison). Moreover, multiplying a CSP value by the mass of the earth ($6 \times 10^{24}$ kg) is not a meaningless exercise. The maximum value of the CSP has been found to be surprisingly constant for all cases other than those containing a large earthquake. Of the five cases (different days, all at the same starting time of a 12-h record) shown in Fig. 8, only the black curve departs significantly from a maximum at 6000 s of about -16.3. A single seismometer at one site cannot adequately estimate the total vibrational power of the earth; however, an average over the maximum CSP times earth mass of a large number of instruments situated geographically such as to `cover' the earth should yield a decent estimate of the average vibrational power in watts of the earth.

**Secular trends in the maximum CSP**
The maximum values of the CSP, based on records of the same type as Fig. 8, were used to generate Fig. 9. The data in all cases were from a VolksMeter located in Redwood City, CA. The remarkable feature of this graph is the secular increase that is evident for reason of the relatively small diurnal fluctuations in the maximum value.

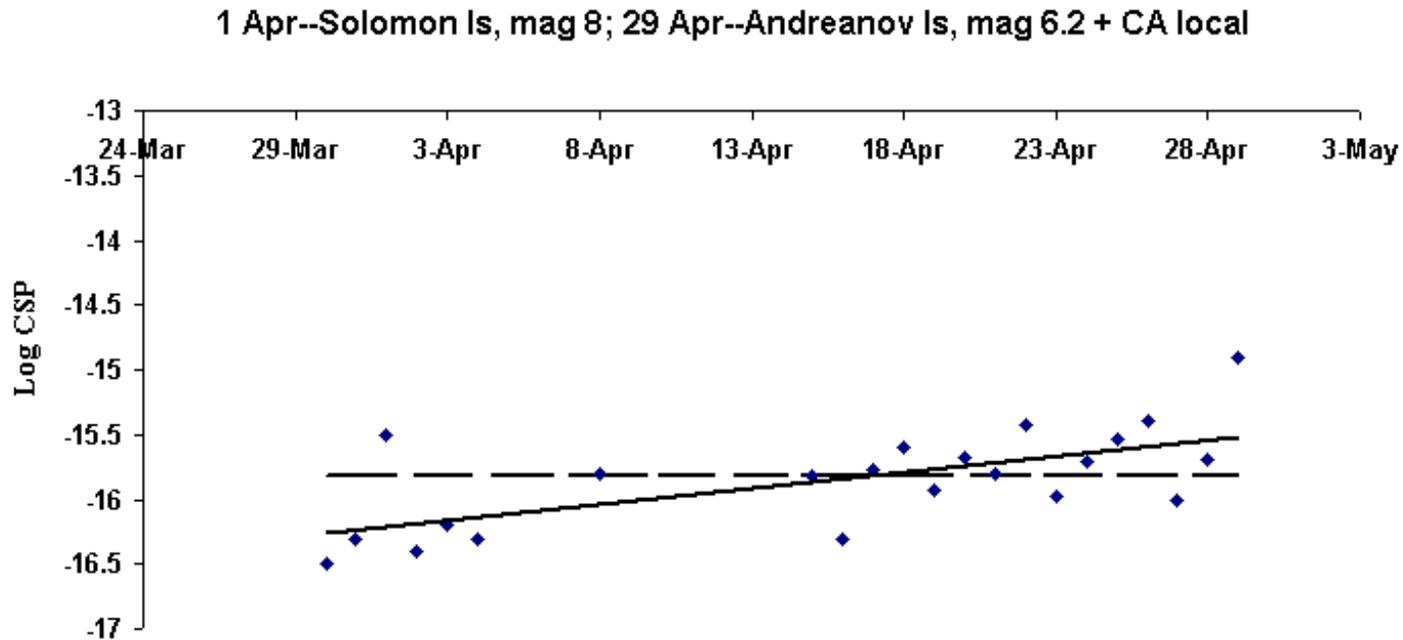

**Figure 9.** Graph of the maximum value of the CSP over a one-month inverval. Two teleseisms and one smaller, local earthquake occurred during this time. The solid line is a trendline fit to the data and the dashed line is the mean value of the data set.

**Acknowledgment** John Lahr, retired USGS seismologist, has provided valuable input during the development of this article. His recommendations have been very beneficial and greatly appreciated.

BIBLIOGRAPHY

[1] J. Berger & P. Davis, G. Geophys. Res., Vol. 109, B11307 (2004).

[2] The VolksMeter is sold by RLL Instruments, online information at http://rllinstruments.com

[3] D. McNamara & R. Buland, ``Ambient noise levels in the continental United States, manuscript in review: BSSA, online article at http://geohazards.cr.usgs.gov/staffweb/mcnamara/PDFweb/McNamaraBuland.pdf.

[4] Wikipedia definition of PSD, online at http://en.wikipedia.org/wiki/Power_spectral_density

[5] R. Peters articles: (i) ``Fourier transform construction by vector graphics'', Am. J. Phys. 60, 439 (1992), and (ii) ``Graphical explanation for the speed of the Fast Fourier Transform'', online link at http://web.mit.edu/redingtn/www/netadv/XfastFouTr.html

[6] R. Peters, Two Damping chapters for *Vibration and Shock Handbook''*, C. de Silva ed, CRC, ISBN 0-8493-1580-8 (2005); also, online article ``Oscillator damping with more than one mechanism of internal friction damping'', http://arxiv.org/html/physics/0302003/.

[7] R. Peters, ``Friction at the mesoscale'', Contemporary Physics, Vol. 45, No. 6, 475-490 (2004).

[8] slide no. 20, R. Peters, AGU presentation, ``State of the art digital seismograph'', online link at http://seismicnet.com.

---

File translated from T$_E$X by T$_T$H, version 1.95.

On 8 May 2007, 09:22.